\documentclass[12pt,preprint]{aastex}
\usepackage{graphicx}

\begin{document}
\title{On the Completeness of Reflex Astrometry on Extrasolar Planets near the Sensitivity Limit}
\shorttitle{Reflex Astrometry}
\author{Robert A. Brown}
\affil{Space Telescope Science Institute, 3700 San Martin Drive, Baltimore, MD 21218}
\email{rbrown@stsci.edu}

\begin{abstract}
We provide a preliminary estimate of the performance of reflex astrometry on Earth-like planets in the habitable zones of nearby stars. In Monte Carlo experiments, we analyze large samples of astrometric data sets with low to moderate signal-to-noise ratios. We treat the idealized case of a single planet orbiting a single star, and assume there are no non-Keplerian complications or uncertainties.  The real case can only be more difficult. We use periodograms for discovery and least-squares fits for estimating the Keplerian parameters. We find a completeness for detection compatible with estimates in the literature. We find mass estimation by least squares to be biased, as has been found for noisy radial-velocity data sets; this bias degrades the completeness of accurate mass estimation. When we compare the true planetary position with the position predicted from the fitted orbital parameters, at future times, we find low completeness for an accuracy goal of 0.3 times the semimajor axis of the planet, even with no delay following the end of astrometric observations. Our findings suggest that the recommendation of the ExoPlanet Task Force (Lunine et~al.\ 2008) for ``the capability to measure convincingly wobble semi-amplitudes down to 0.2~$\mu$as integrated over the mission lifetime,'' may not be satisfied by an instrument characterized by the noise floor of the \textit{Space Interferometry Mission}, $\sigma _\mathrm{floor} \approx 0.035~\mu$as. An important, unsolved, strategic challenge for the exoplanetary science program is figuring out how to predict the future position of an Earth-like planet with accuracy sufficient to ensure the efficiency and success of the science operations for follow-on spectroscopy, which would search for biologically significant molecules in the atmosphere.
\end{abstract}


\section{INTRODUCTION}

Today, the question of life in the universe is a compelling goal of scientific research. Finding such life would have vast implications for the human mind, consequences that seem real to everyone, even if difficult to express. Down through the ages, the uniqueness of Earth has been a stimulating issue for philosophers and ordinary people alike. Today, inspired by the discovery of many large planets around nearby stars, and propelled by progress in utilizing broad wavefronts of astronomical light, we can at last prepare to test the Earth's exceptionalism by means of large optical systems in space. 

An important, specific goal is to search for evidence of biologically significant molecules, particularly free oxygen, in spectra of the atmospheres of Earth-size planets located in the habitable zones of nearby stars---planets with likely surface temperatures in the range compatible with water occurring in liquid form. Recently, the ExoPlanet Task Force of the Astronomy and Astrophysics Advisory Committee (Lunine et~al.\ 2008) recommended---as a precursor to such direct detection and study---a space astrometry mission with ``the capability to measure convincingly wobble semi-amplitudes down to 0.2~$\mu$as integrated over the mission lifetime,'' where the astrometric semi-amplitude for zero eccentricity is
\begin{equation}
\label{eq1}
\alpha =3\frac{m_p a}{m_s D}~\mu\mathrm{as}~~.
\end{equation}
Here, the range of $a$, which is the planetary semimajor axis in AU, is the habitable zone, or 0.7--1.5 AU for a star of $L_s=1.0$ solar luminosity, and otherwise varying as $\sqrt{L_s}$; $m_{p}$ is the planetary mass in Earth units; $m_{s}$ is the stellar mass in solar units; and $D$ is the distance of the star in parsecs. For the Earth-Sun system observed at $D = 10$~pc, $\alpha = 0.3~\mu$as.

There are several good reasons for the ExoPTF's recommendation on space astrometry. Nevertheless, there has been no detailed explanation of what the numerical goal ``0.2~$\mu$as'' implies, either in terms of expected scientific performance or regarding the true requirements on a mission to achieve it. Is it a good number? Is it achievable by any currently planned astrometric mission? These questions are sharpened by the bias in least-squares estimation of Keplerian parameters near the sensitivity limit, most recently discussed by Shen \& Turner (2008) for radial-velocity studies, and documented here for astrometry. Therefore, the immediate purpose of this paper is to expose some performance issues that attend reflex astrometry with noisy data using the least-squares estimator in most common use.  (Alternative estimators might be less biased or unbiased. Bayesian estimation is identical to least-squares when the errors are normally distributed and the priors are flat, as in the experiments described here.) The ultimate goal is to find, if possible, a coherent program of space astrometry and follow-on coronagraphic spectroscopy of Earth-like planets that reasonably fulfills the expectations of the community, as well as the public.

In the first instance, our expectations are that astrometry can overcome three technical difficulties that have been identified with the direct-detection approach: (1)~low search completeness per observation, implying many long exposures with no results, (2)~low final search completeness for many high-priority stars, due to a high fraction of permanently obscured habitable-zone orbits, and (3)~low probability of recovering in the future any planets that are discovered, because of the difficulty in estimating the planetary orbit from the limited range of the few direct astrometric measurements that might be obtained in the first observing season---the season of discovery---before the target is lost in the Sun (Brown 2005; Brown, Shaklan, \& Hunyadi 2007; Pravdo et~al.\ 2007).

Of no less importance, at least qualitatively, our expectations are also scientific, namely that astrometry will provide important physical information that direct detection cannot provide, particularly estimates of the planetary mass, which controls atmospheric structure and composition, and orbit size, which determines the planet's thermal regime. Such information is basic for understanding the planet's physical conditions and, indeed, its very identity. It may also provide clues about a planet's origins and dynamical evolution.

Planets cannot hide from astrometry, except in the noise. In principle, only astrometry can measure the true mass and determine all the orbital elements. In principle, only astrometry can ensure that follow-on spectroscopy is planned with accurate predictions of the separation, position angle, and brightness of the planet at future times.  \textit{In practice}, however, as we show here using the completeness metric, these expectations may not be satisfied by the design parameters currently associated with the observing system most strongly advocated for the task, the \textit{Space Astrometry Mission}, light version, or \textit{SIM Lite}. 

\section{THE BASICS}

We assume an idealized astrometric instrument, which measures the two-dimensional position of the star relative to a local astrometric reference frame on the sky.  An idealized astrometric data set comprises $N$~measurements of the form ($t_i$, $\tau_i$, $x_i$, $y_i$, $\sigma_i$): the $x$--$y$ position at epochs $t_{i}$ spread over duration $T$, longer than one planetary period, where the exposure time is $0.5\,\tau_{i}$ for each of the two quasi-orthogonal directions, and where $\sigma_i =\sigma_0 (\tau_0/\tau_i)^{1/2}$ is the positional uncertainty in either direction.  We assume the astrometric errors are normally distributed. 

Assuming that $D$ is known from the annual parallax and that $m_{s}$ is known from stellar spectrophotometry, the theoretical apparent position of the star is fully described by seven Keplerian parameters, of which three are physical properties of the system ($a$, eccentricity $e$, and $m_{p}$), three are Eulerian rotations of the orbit with respect to the line of sight (argument of periastron $\omega$, inclination angle $i$, and longitude of the ascending node $\Omega$), and one specifies the orbital phase at a moment in time, such as at  the epoch of the first data point (initial mean anomaly $M_{0}$). Currently, we consider only planetary orbits with periods significantly shorter than $T$.

We consider a subset of the universe of astrometric reflex signals due to planets on randomized habitable-zone orbits for which $a$ is uniformly distributed in the habitable-zone range $0.7\sqrt{L_s} \le a\le 1.5\sqrt{L_s}$, $e$ is uniformly distributed over 0 $\le e \le 0.35$; and $\omega$, $i,$ and $\Omega$ are randomized to distribute the orbit pole uniformly on the sphere. The subset of signals we consider is made up of those for which a key characteristic, the astrometric signal-to-noise ratio \textit{SNR}, 
\begin{equation}
\label{eq2}
\mathit{SNR}\equiv\alpha/\sigma~~,
\end{equation}
is sufficiently small to approach the limits of useful scientific performance.  Here, the ``mission error" for a star is
\begin{equation}
\label{eq3}
\sigma \equiv \sigma_0 \tau_0^{1/2} 
\left( {\sum\nolimits_{i=1}^N {\tau _i } } \right)^{-1/2}
=\sigma_0 \tau_0^{1/2} \tau^{-1/2}~~,
\end{equation}
where $\tau$ is the total exposure time. The noise reduction in Eq.~3 with $\tau$ and $N$ is valid only down to the noise floor, or for $\sigma>\sigma_\mathrm{floor}$, where $\sigma_\mathrm{floor}$ is determined by systematic effects. 

\textit{SNR} can be small due to any combination of the several instrumental, observational, and physical parameters appearing in Eqs.~1, 2, and 3. Nevertheless, our interest is focused on $\alpha \approx 0.2~\mu$as, which is the ExoPTF's recommended goal, as well as 2/3 the iconic value for the Earth-Sun system at $D = 10$~pc. (Due to orbital eccentricity and the arbitrary alignment of the orbital major axis with respect to the line of sight, $\alpha$ may be larger or smaller than the semi-major axis of the apparent orbital ellipse on the plane of the sky.)

We use the completeness metric $C$(\textit{SNR}) to estimate scientific performance. We define $C$ as the fraction of possible planets with the characteristic \textit{SNR} that are successfully observed, assuming the planet is present. ``Successfully observed'' variously means (1)~that the planet is detected by periodogram analysis of the data, or if it is detected, (2)~that the planetary mass is, or (3)~future position is, usefully estimated from the orbital solution produced by a least-squares fit of the parameters of the Keplerian theory to the noisy data set.

\section{PREVIOUS PERFORMANCE ESTIMATES FOR \emph{SIM LITE}}

Traub et~al.\ (2009) is the most recent discussion of the scientific performance of space reflex astrometry on Earth-like planets, assuming the design parameters of \textit{SIM Lite}. That paper reports the preliminary findings of a JPL-sponsored study in which one ``goal was to see what accuracy of \textit{SIM Lite} is needed to detect Earth-like planets.'' The study builds on earlier analyses by the \textit{SIM} science team of variants of the \textit{SIM} mission, including Catanzarite et~al.\ (2006) and Unwin et~al.\ (2008). The findings have been reiterated by the astrometry committee of the JPL-sponsored 2008 Exoplanet Forum (Lawson, Traub, \& Unwin 2009;  Muterspaugh \& Tanner 2009). One purpose of that forum was to inform the Decadal Survey of Astronomy and Astrophysics (Astro2010) ``concerning technology development and technology readiness of proposed exoplanet missions,'' in this case, \textit{SIM} and its variants. The forum report urges that a \textit{SIM}-type astrometry mission be implemented forthwith. It cites the ExoPTF finding ``that for the 11--15 year (2019--2023) timeframe, `Assuming the space-borne astrometric mission described above is fielded in the second time epoch, no additional major space-based astrometric effort is envisioned in this time frame.' The committee agrees with this conclusion.''

We infer the following common denominators of the studies and reports 
mentioned in the previous paragraph.
\begin{enumerate}
\item  For \textit{SIM Lite}, the instrumental parameters are $\sigma_{0}=1.41~\mu$as, $\tau_{0}=2200$~sec, and $\sigma_\mathrm{floor}=0.035~\mu$as, which from Eq.~3 implies $N_\mathrm{max}=1600$ and $\tau_\mathrm{max}=3.6\times10^6$~sec. For $\alpha=0.2~\mu$as (ExoPTF) or $\alpha=0.3~\mu$as (Earth-Sun at 10~pc), $\textit{SNR}_\mathrm{max}=5.7$ or 8. We accept these numbers as given.

\item For detection via periodogram (power spectrum), \textit{SNR}~= 5.8 is required for search completeness $C = 0.5$ with a false-alarm probability \textit{fap}~= 0.01. We agree with this, finding the value is closer to \textit{SNR}~= 6.0 for our adopted planetary population, as shown below.  In any case, the \textit{SIM Lite} search completeness for the ExoPTF's $\alpha=0.2~\mu$as is $C\approx0.5$.

\item Estimates of planetary mass are achievable with accuracy ``close to the theoretically expected value'' from the minimum-variance bound or Cram\'{e}r-Rao limit (Traub et~al.\ 2009, Gould 2008). Using the least-squares estimator, we differ, but actually it depends on the meaning of ``close.'' We show below that, because the least-squares estimator of mass is biased, the completeness 
$C$ of even roughly accurate estimates of mass for Earth-like planets is significantly degraded  in the \textit{SIM Lite} domain of $\sigma >\sigma_\mathrm{floor}\approx0.035~\mu$as.

\item The estimates of the orbital period approach the minimum-variance bound. 
We agree. Just as Shen \& Turner (2008) found for radial-velocity data sets, for example, we find no sign of bias for period estimation via least-squares fitting of noisy astrometric data sets.

\item Other than our studies reported below, we are not aware of any previous tests of the ability to accurately predict future planetary positions from the analysis of noisy astrometric data sets. Nevertheless, the literature contains fulsome assertions about the adequacy of \textit{SIM}-based positional predictions, even if their basis has not been explained. Unwin et~al.\ (2007): ``For many of these stars, \textit{SIM}'s orbital solution will be precise enough to predict the best timing for a direct observation. This information is crucial for direct imaging, since a planet in the habitable zone can spend much of its time hidden in the glare of the planet star.''  Lunine et~al.\ (2008): ``Because the astrometric mission would have already determined the locations (`addresses') and orbits of the candidate planets, the spaceborne direct detection system is relieved of the burden of searching; its mission is to point at those stars and study the planets. This greatly simplifies a direct detection mission in several ways: {\ldots}The target list can be ordered and prioritized ahead of time to maximize efficiency in moving from one target to another and in observing those targets whose orbital phases make them most observable{\ldots}'' Muterspaugh \& Tanner (2009): ``Astrometric observations of stars hosting Earth-like planets allow the orbits to be determined. The position of the planet can be predicted as a function of time. This provides a solid basis for future planning [of] direct imaging programs, as the observing schedule can be set to look at the targets when the star and planet are most optimally configured for isolating light from the planet.''  In the current study, we do not confirm these assertions using the least-squares estimator of the Keplerian parameters. In the \textit{SIM Lite} performance domain, we find low $C$ for even roughly accurate predictions of the future positions of Earth-like planets.
\end{enumerate}

\section{DISCOVERY VIA PERIODOGRAM}

Currently, the most popular algorithm to search noisy data for faint periodic signals is the periodogram, $Z(f)$, where $f$ is the angular frequency. In this paper, we use the most popular normalization of the periodogram, which subtracts the mean value of the measurements and divides by the standard deviation of the data set (Lomb 1976; Scargle 1982; Horne \& Baliunas 1986). Recently, the theoretical basis for closed-form estimates of periodogram performance has collapsed spectacularly (Frescura, Engelbrecht, \& Frank 2008). A new calibration approach, based wholly on Monte Carlo methods, is now emerging (Catanzarite et~al.\ 2006), and is followed here. 

For two-dimensional astrometry, we use the ``joint'' periodogram defined by Catanzarite et~al.\ (2006). 

Instead of selecting the maximum peak value $Z_\mathrm{max}$ at a finite list of ``independent frequencies''---a now-discredited theoretical concept, 
particularly for arbitrarily spaced data---we locate the true peak in the 
continuous range of frequencies between the fundamental, 2$\pi/T$, and the 
Nyquist frequency, $\pi N/T$. (Orbital frequencies in the habitable zone are always near the short end of this range.)

At its performance limits, discovery via periodogram is based on selecting a 
value of \textit{fap}, then estimating the threshold value for the maximum peak, $Z_\mathrm{threshold}$, of $Z(f)$ such that data sets of pure noise---with no periodic signal present---produce $Z_\mathrm{max~peak} \ge Z_\mathrm{threshold}$ with a probability equal to \textit{fap}. The value of $Z_\mathrm{threshold}$ depends on $N$, but not on the timing of the data points. For all the periodogram analysis report here, we adopted \textit{fap}~= 0.01, and we estimated the values of $Z_\mathrm{threshold}$ for given $N$ from 100,000 data sets of pure noise.

For discovery, we are interested in the search completeness $C(\textit{SNR})$, which is the fraction of random planets with given \textit{SNR} that are detected by the criterion $Z_\mathrm{max~peak} \ge Z_\mathrm{threshold}$. For this purpose, we prepared 880,000  random data sets at spaced values of \textit{SNR} in the range $5\le\textit{SNR}\le100$, then subjected the data sets to periodogram analysis. The results are shown in Fig.~1. We confirm that $C = 0.5$ is achieved at \textit{SNR} $\approx6$. This result---indeed the value of $C$ at any \textit{SNR}---depends somewhat on the assumed planetary population. For example, face-on orbits achieve the highest $C$ at any \textit{SNR}, and high eccentricity degrades $C$, due to the spectral power distributed into harmonics.

The shaded region of Fig.~1 is the range of \textit{SNR} that is excluded for \textit{SIM Lite} on Earth-like planets at 10~pc, $\textit{SNR}_\mathrm{max}\approx8$. At \textit{SNR}$_\mathrm{max}$, $C = 0.85$ for discovery.

In the process of conducting these periodogram investigations, we confirmed 
that search completeness $C$ depends on the parameters $\alpha$, $\sigma$, $\tau$, and $N$ only as they are combined in the \textit{SNR} characteristic according to Eqs.~2 and~3, as expected from the literature. 

\section{MASS ESTIMATION}

At spaced values of \textit{SNR} in the range $5\le\textit{SNR}\le100$, we found orbital solutions for 265,000 data sets in which periodogram analysis had discovered a planet. We compared the true value of the planetary mass, $m_p$(true)---the value with which the data set was prepared---with the estimated value, $m_p$(est), found by a least-squares orbital fit to the data. In this application, $C$(\textit{SNR}) means the fraction of data sets with the value of \textit{SNR} for which the fractional error, ($m_p$(est)--$m_p$(true))/$ m_p$(true), occurs in a specified range centered on zero. Fig.~2 shows the results for the ranges $\pm$10\% and $\pm$25\%. We see that for the mass-estimation accuracy $\pm$25\%, $C < 0.70$ for $\textit{SNR}<8$, and for $\pm$10\% accuracy, $C < 0.35$. 

The dashed lines in Fig.~2 show the performance associated with the 
minimum-variance bound (Gould 2008). At this bound, the expected fractional 
error in mass is $\sqrt 2/$\textit{SNR}, and the normalized deviations---observed 
fractional errors in mass times \textit{SNR}$/\sqrt 2$---should be normally 
distributed with mean zero and unit variance (Traub \& Gould 2008). Therefore, for mass estimation accuracy $q$, which equals 0.1 or 0.25 for the cases shown, the completeness associated with the minimum-variance bound is
\begin{equation}
\label{eq3}
C(q;\textit{SNR})=\textit{erf}\left(\frac{q~\textit{SNR}}{2}\right)~~,
\end{equation}
where \textit{erf} is the error function. Comparing the dashed and solid curves, it is clear that the performance estimated by these Monte Carlo simulations, using least-squares estimation, falls significantly short of that expected from the minimum-variance bound for $\textit{SNR}<8$.

A closer look at the normalized deviations (Fig.~3) reveals the reason for the reduction of $C$ for mass estimation from the minimum-variance bound: the 
least-squares estimator produces biased estimates of mass from noisy astrometric data sets, just as Shen and Turner (2008) found for noisy radial-velocity data sets. 

In the process of conducting these mass investigations, we confirmed that $C$ for accurate mass estimation depends on the parameters $\alpha$, $\sigma$, $\tau$, and $N$ only as they are combined in \textit{SNR} according to Eqs.~2 and~3.

\section{PREDICTIONS OF THE PLANETARY POSITION AT FUTURE TIMES}

Habitable-zone orbits of nearby stars are highly obscured, with many possible planets peeking out only for short times from behind a central field obscuration. Also, suppression of starlight may be easier to achieve in a smaller region of the focal plane. In response to these considerations, we suggest that a positional accuracy of 30\% the semi-major axis---$\pm0.3~a$---is a reasonable first estimate of a possible requirement on astrometry to support the planning of direct observations of the planet. 

We analyzed the same sample of 265,000 pairs of true and fitted orbital solutions 
discussed in the previous section to estimate performance in predicting the 
planetary position. The six solid curves in Fig.~4 show $C$(\textit{SNR}) for the positional metric better than $\pm$0.3~$a$ after 0--20~years have elapsed after the end of the 5-year duration of astrometric observations.  The positional metric is defined as the norm of the vector deviation between the true and estimated planetary positions on the plane of the sky. For $\textit{SNR}<8$, we find $C < 0.5$ after more than one year has elapsed, and $C < 0.2$ after ten~years.
\newpage

For a further, more elaborate, metric of recoverability, we considered the time 
interval of one (estimated) orbital period following the end of astrometric observations. We computed the true position of the planet for each day from the true orbital elements, and computed the estimated position from the estimated orbital elements. Next, we identified the subset of days for which the planet was predicted (from the estimated elements) to have a greater separation from the star than 0.7~$a$, which seems like a reasonable definition of \textit{when} follow-up spectroscopy should be planned. Next, we checked on those days 
whether the norm of the deviation between the true and estimated positions of the planet was less than 0.3~$a$, which seems like a reasonable working size of the dark hole that a coronagraph might create around the entrance aperture of a spectrometer. At each \textit{SNR}, we counted the number of days for which both these criteria were satisfied, and divided it by the number of days only the first was met. This recoverability metric tends to unity in the limit of excellent  predictions, at high \textit{SNR}. If the metric is $>$\,0.95, say, it means that the planet is recovered (within 0.3~$a$ of the estimated location) for $>$\,95\% of the days that the data predicts the planet to be observable (outside 0.7~$a$). Preliminarily, we define a planet as ``recovered" if the metric is $>$\,0.95. The dashed red line in Fig.~4 shows $C$(\textit{SNR}), the recovery completeness by this test, $C < 0.1$ for $\textit{SNR}<8$.

In the process of conducting these positional investigations, we confirmed that $C$ depends on the parameters $a$, $\sigma$, $\tau$, and $N$ only as they are combined in the \textit{SNR} characteristic according to Eq.~2.

\section{DISCUSSION}

Our preliminary analysis suggests that the scientific performance on Earth-like planets of a space astrometry mission characterized by $\sigma_\mathrm{floor} \approx 0.035~\mu$as may not satisfy expectations, particularly, the ExoPTF recommendation for ``the capability to measure convincingly wobble semi-amplitudes down to 0.2~$\mu$as integrated over the mission lifetime,'' particularly in estimating the planetary mass and future planetary position.  The reason for this underperformance is that the maximum astrometric signal-to-noise ratio, $\textit{SNR}_\mathrm{max}=\alpha/\tau_\mathrm{floor}=5.7$ for $\alpha=0.2~\mu$as, which is the \textit{SNR} required for $\sim$50\% discovery completeness, is well below the \textit{SNR}s required for high completeness of accurate estimates of planetary mass and future positions.

In the case of mass estimation, the degradation from the minimum-variance 
bound is only quantitative, and perhaps not a serious problem for science. For example, we expect that meaningful and useful \textit{upper limits} to the planetary mass can still be set by reflex astrometry, as is customary for the radial-velocity technique. 

In the case of predicting the planetary position, the problem is qualitative and  more serious. Knowing where and when to place the entrance aperture of the spectrometer is a huge science-operational issue for the follow-on mission, due to the very long exposure times. We had hoped that \textit{SIM}-class astrometry would be a solution, but now that appears not to be the case, at least for the most interesting planets in terms of life, which are the Earth-like planets at typical distances of 5--20~pc. 

The bias in least-squares estimation of the Keplerian parameters certainly 
contributes to the error in positional predictions.

This paper reports only a preliminary analysis, not yet taking into account the characteristics of individual stars, which affect both astrometric performance and follow-up coronagraphic spectroscopy. Also, we treat the idealized case of a single planet orbiting a single star, and assume there are no non-Keplerian complications or uncertainties---such as due to multiplicity, proper motion, parallax, stellar activity, reference-frame errors, etc.  The real case can only be more difficult.

The findings presented here are a source of concern about the true answer to the question posed by Traub et~al.\ (2008): ``What accuracy of \textit{SIM Lite} is needed to detect Earth-like planets?'' 

We hope and expect that these results will motivate a critical reexamination of  exoplanetary program strategies involving space astrometry.  We must ensure either that an adequate technical solution for reflex astrometry is part of the plan, or we must define a programmatic pathway toward spectroscopy on Earth-like planets that is less dependent on reflex astrometry.

An important unsolved, strategic problem in the exoplanetary science 
program is figuring out how to predict the future position of an Earth-like planet with accuracy sufficient to ensure the success of the science operations of a follow-on spectroscopic mission, which would search for biologically significant molecules in the atmosphere.

\begin{acknowledgements}
I appreciate Jeremy Kasdin's raising the issue of the accuracy of the planetary ephemeris, and R\'emi Soummer's calling this to my attention.
I thank Christopher Burrows for interesting discussions, both directly and indirectly related to this work. I thank Christian Lallo, for making the computations possible, and Sharon Toolan for her excellent help preparing the manuscript.
\end{acknowledgements}

\begin{figure}
\plotone{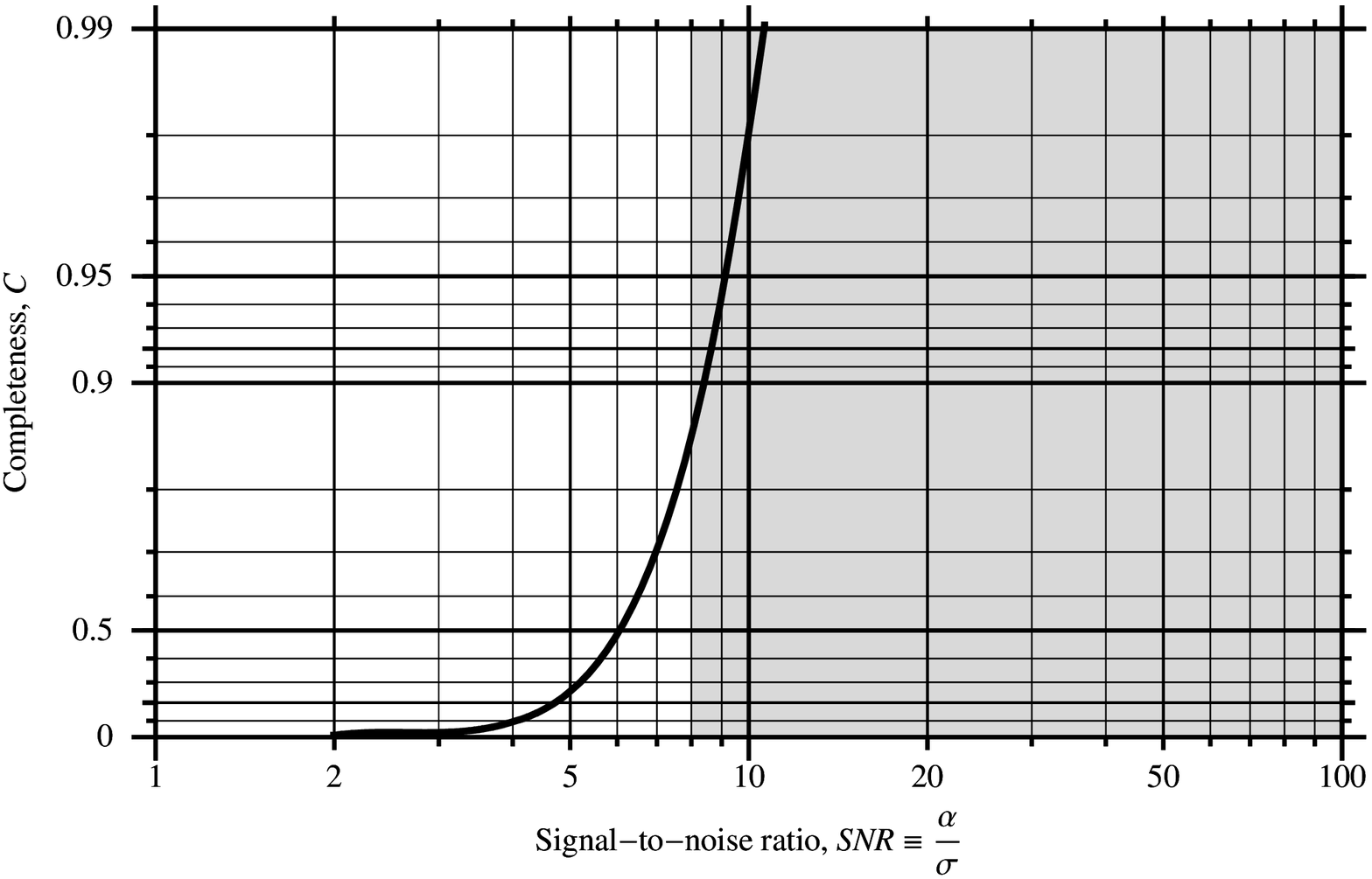}
\label{fig1}
\caption{Discovery search completeness of reflex astronometry as a function of  \textit{SNR}. $C$ is the fraction of possible planets drawn from the assumed population that are detected by periodogram with $\textit{fap}=0.01$. The shaded region is denied to \textit{SIM Lite} ($\tau_\mathrm{floor}=0.035~\mu$as) for $\alpha<0.3~\mu$as, which is the reflex semi-amplitude of the Earth-Sun system observed from $D=10$~pc.}
\end{figure}

\begin{figure}
\plotone{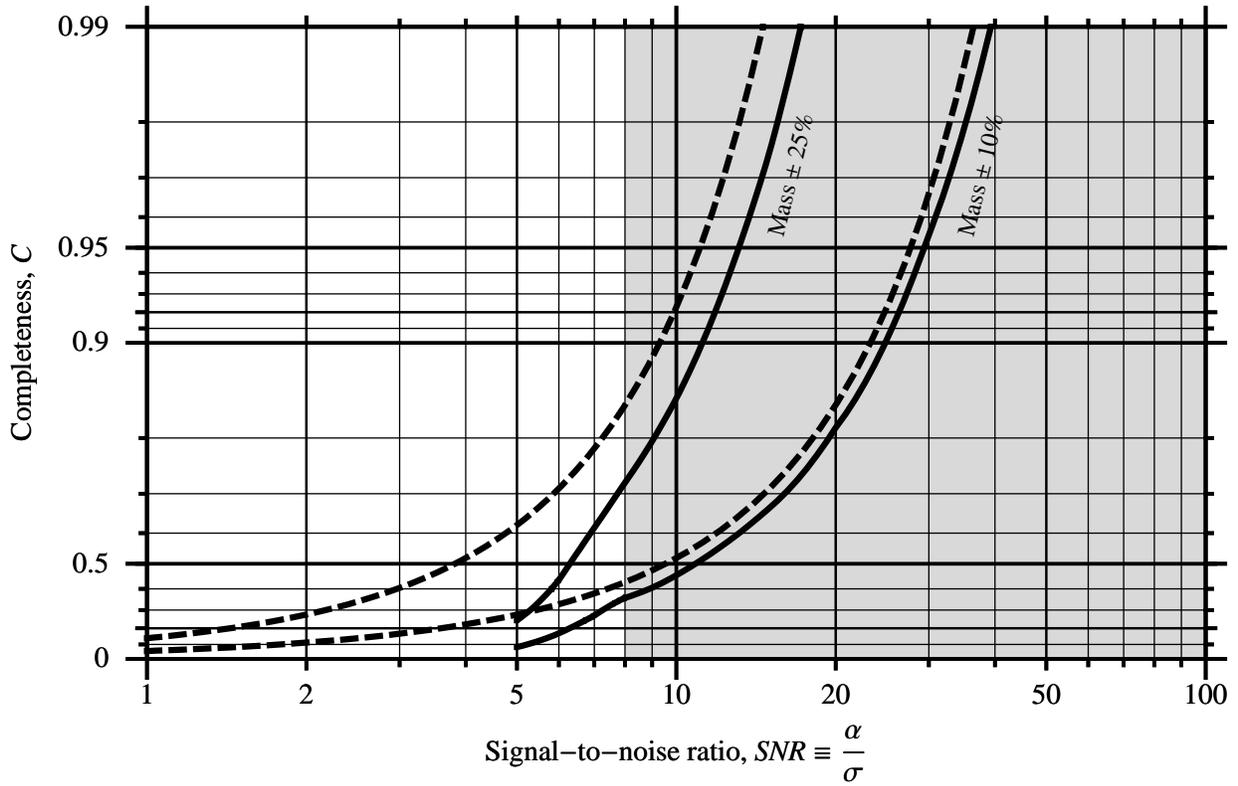}
\label{fig2}
\caption{Completeness of mass estimation with $\pm$10\% and $\pm$25\% accuracy.  Solid: results of Monte Carlo experiments in which orbital solutions for 265,000 noisy data sets were obtained by least-squares fits. Dashed: the predicted completeness if mass estimation achieved the minimum-variance bound---which it does not here, due to bias in the least-square estimator.}
\end{figure}

\begin{figure}
\plotone{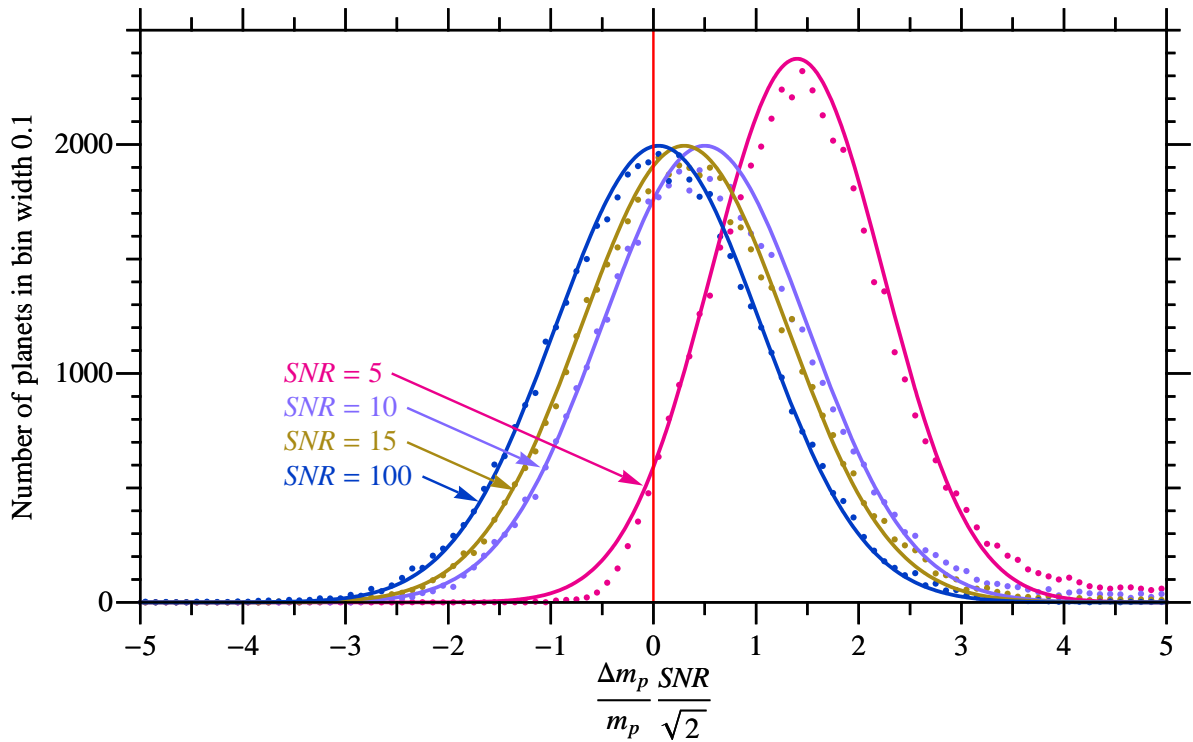}
\label{fig3}
\caption{Bias in estimating $m_p$ by least-squares fit of Keplerian 
parameters to astrometric data sets. Histogram of the fractional deviation 
of the estimated mass from the true mass multiplied by \textit{SNR}$/\sqrt 2$, which should produce a normal distribution with zero mean and unit variance for estimation satisfying the minimum-variance bound (Traub \& Gould 2008). The indicated bias accounts for degradation in $C$ for accurate mass determination, as indicated by the separation of the solid and dashed curves in Fig.~2.}
\end{figure}

\begin{figure}
\plotone{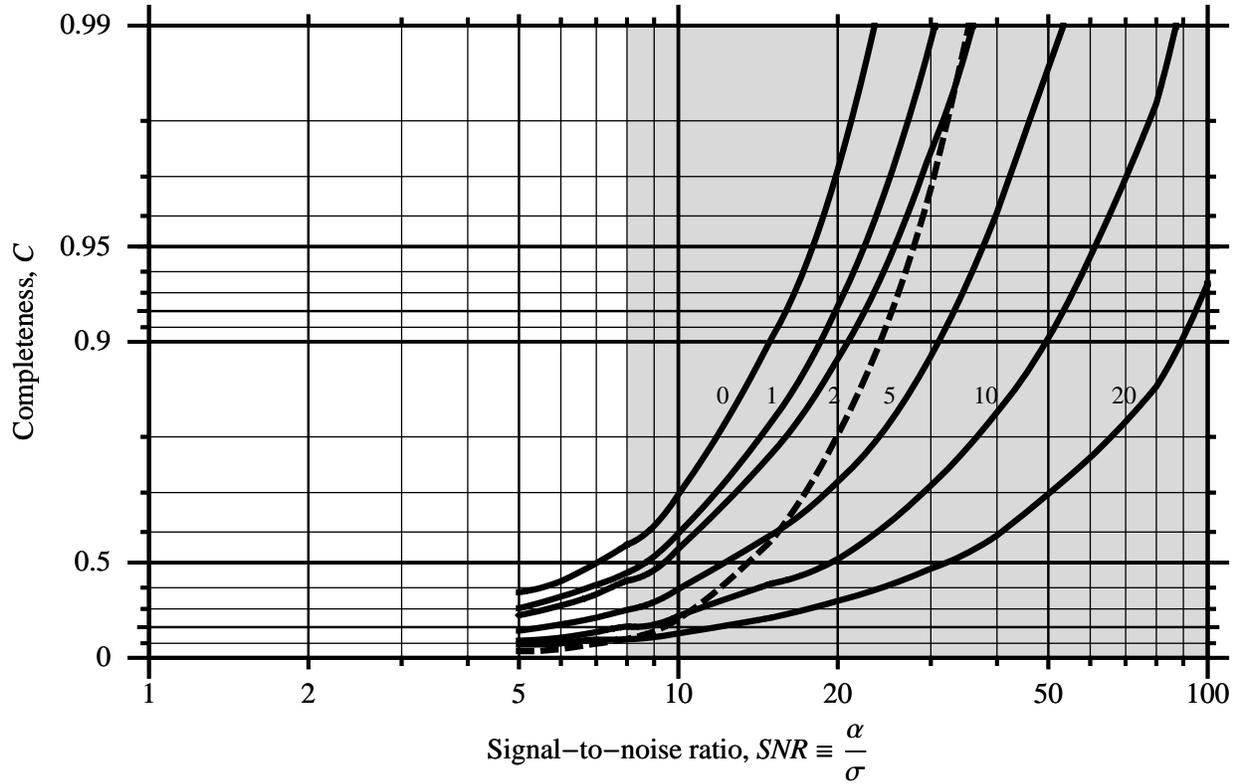}
\label{fig4}
\caption{Various completeness measures of the accuracy of predicting the future planetary position from orbital solutions obtained by least-squares fits to noisy astrometric data sets.  Solid: the completeness of positional estimates within $\pm3~a$ for 0--20 years delay, as indicated, after the 5-year duration of observations.  Dashed: results of a recovery metric test described in the text.}
\end{figure}
\end{document}